\documentclass[aps,prd,floatfix,superscriptaddress,preprintnumbers,nofootinbib,twocolumn,longbibliography]{revtex4-1}
\pdfoutput=1

\usepackage{amsmath, amssymb, amsfonts, graphicx, mathrsfs, verbatim, float, slashed, bbm, color, xcolor, xspace, enumerate, url, bbding, multirow, outlines,upgreek}
\usepackage[english]{babel}
\usepackage[colorlinks=true,linkcolor=black,citecolor=darklightsabergreen]{hyperref}
\definecolor{darklightsabergreen}{rgb}{0.0, .49, 0.06}

\newcommand{\acro}[1]{\textsc{\MakeLowercase{#1}}} 

\newcommand{\gsim}{\gtrsim}
\newcommand{\lsim}{\lesssim}
\newcommand{\ra}{\rightarrow}

\def\Oc{\mathcal{O}}

\newcommand{\beq}{\begin{equation}}
\newcommand{\eeq}{\end{equation}}
\newcommand{\bea}{\begin{eqnarray}}
\newcommand{\eea}{\end{eqnarray}}
\newcommand{\nn}{\nonumber}
\newcommand{\1}[1]{\, \mathrm{#1}} 

\def\mx{m_\chi}
\def\mdm{m_\chi}
\def\mDM{m_\chi}
\def\rhodm{\rho_\chi}

\def\texp{t_{\rm exp}}

\def\LD{L_{\rm det}}
\def\AD{A_{\rm det}}
\def\ntarget{n_{\rm det}}
\def\mtarget{m_{\rm T}}

\def\vhalo{v_\chi}
\def\sigmaNDM{\sigma_{{\rm n} \chi}}
\def\sigmaT{\sigma_{{\rm T}\chi }}
\def\sigmathresh{\sigma_{\rm MIMP}}

\def\sigmasos{\sigma_{\rm sos}}

\def\mDMmax{m_\chi^{\rm max}}

\def\muTDM{\mu_{{\rm T}\chi}}
\def\munDM{\mu_{{\rm n}\chi}}
\definecolor{darklightsabergreen}{rgb}{0.0, .49, 0.06}
\definecolor{orange}{rgb}{1.0, 0.5, 0.0}

\begin{document}

\title{Saturated Overburden Scattering and the Multiscatter Frontier: \\ Discovering Dark Matter at the Planck Mass and Beyond}

\author{Joseph Bramante}
\affiliation{CPARC and Department of Physics, Engineering Physics, and Astronomy, Queen's University,  Kingston, Ontario, K7L 2S8, Canada}
\affiliation{Perimeter Institute for Theoretical Physics, Waterloo, Ontario, N2L 2Y5, Canada}

\author{Benjamin Broerman}
\affiliation{CPARC and Department of Physics, Engineering Physics, and Astronomy, Queen's University,  Kingston, Ontario, K7L 2S8, Canada}

\author{Rafael F. Lang}
\affiliation{Department of Physics and Astronomy, Purdue University, West Lafayette, IN 47907, USA}

\author{Nirmal Raj}
\affiliation{Department of Physics, University of Notre Dame, Notre Dame, IN 46556, USA}

\begin{abstract}
We show that underground experiments like \acro{LUX}/\acro{LZ}, \acro{PandaX-II}, \acro{XENON}, and \acro{PICO} could discover dark matter up to the Planck mass and beyond, with new searches for dark matter that scatters multiple times in these detectors. This opens up significant discovery potential via re-analysis of existing and future data. 
We also identify a new effect which substantially enhances experimental sensitivity to large dark matter scattering cross-sections: 
while passing through atmospheric or solid overburden, there is a maximum number of scatters that dark matter undergoes, determined by the total number of scattering sites it passes, such as nuclei and electrons. 
This extends the reach of some published limits and future analyses to arbitrarily large dark matter scattering cross-sections.
\end{abstract}

\maketitle

\section{Introduction}

While the presence of dark matter has been inferred from astrophysical observations and cosmological data, its nature remains enigmatic. 
Dark matter searches in the last few decades have sought out weakly-interacting massive particles, \acro{WIMP}s, which scatter at most once as they pass through underground detectors. 
We show how ongoing underground experiments could find dark matter which scatters multiple times as it travels through these detectors. 
We refer to these multiply interacting particles as \acro{MIMP}s, rather than \acro{WIMP}s. 
We also show that a hitherto-neglected effect in dark matter studies, Saturated Overburden Scattering (\acro{SOS}), enhances published and prospective sensitivities to large dark matter scattering cross-sections.

Dark matter more massive than the so-called unitarity limit of $\sim100~{\rm TeV}$ arises naturally in grand unified theories which predict stable colored and electroweak states~\cite{Burdin:2014xma,Griest:1989wd}.
 It has long been appreciated that super-massive, strongly-interacting, stable particles arise in supersymmetric models~\cite{Raby:1997pb}, and can be produced out of equilibrium in the early universe~\cite{Chung:1998zb,Kuzmin:1998kk,Harigaya:2016vda,Kolb:2017jvz}.
 Depending on the reheating temperature, the equation of state in the primordial epoch, and any substantial increases of entropy in the early universe \cite{Bramante:2017obj,Gelmini:2006pq,Kane:2011ih,Randall:2015xza,Davoudiasl:2015vba}, super-heavy dark matter could make up the bulk of ``missing mass" observed in galaxies. 
For low reheating temperatures, heavy and strongly-interacting dark matter could be a subdominant fraction of mass in the dark sector. Altogether, the discovery of super-heavy relic particles may provide the first evidence for supersymmetry, grand unified theories, and new particle dynamics prior to Big Bang Nucleosynthesis. 
This motivates the search for such particles in the data of existing experiments and with future detectors. 

Here we investigate detection of such heavy, strongly-interacting dark matter. 
In particular, in Section~\ref{sec:sos}, we detail the saturated overburden scattering effect, wherein dark matter scatters with every target along its path through the overburden.
In Section~\ref{sec:multiplicity}, we show the reach obtainable by existing dark matter direct detection experiments looking for multiply interacting dark matter, and evaluate the signatures of such \acro{MIMP}s. 
Section~\ref{sec:directionality} shows how, in much of the \acro{MIMP} parameter space, the angle of entry into the detector can be used to discriminate backgrounds and validate signals, and to determine the mass, cross-section, and local density of \acro{MIMP}s with a single experiment.

\section{Saturated Overburden Scattering}\label{sec:sos}

On its path to a detector, dark matter may be slowed by scattering with the atmosphere, Earth overburden, and detector shielding. 
Usually, as the dark matter-nucleon or dark matter-electron scattering cross-section is increased, the kinetic energy of dark matter arriving at a detector is decreased, due to more frequent scatters en route. 
Because experiments require the dark matter kinetic energy to exceed some minimum energy threshold to observe scattering, dark matter particles arriving too slowly become undetectable.
This results in an upper limit on the cross-section for the sensitivity of such detectors.
However, above a certain cross-section $\sigmasos$, the dark matter scatters with every nucleus or electron along its path. 
Increasing the cross-section beyond  $\sigmasos$ does not further dampen the particle's kinetic energy. 

We thus find that the optical depth $\tau_{\rm od}$, {\em i.e.} the approximate total number of recoils from a single dark matter particle passing through an overburden of length $D$, is correctly given as
\begin{align}
\tau_{\rm od} = \min [ n \sigma D, \; n^{1/3} D]~,
\end{align} 
where $n$ is the number density and $n^{1/3}$ the inverse distance between nuclei or electrons, and $\sigma$ is the dark matter-nucleon or dark matter-electron scattering cross-section. 
The ``min" function accounts for the fact that dark matter scatters at most once with each of the $n^{1/3} D$ nuclei or electrons it passes. 
This treatment is valid so long as the typical momentum transfer is larger than the inverse distance between scattering sites. For super-heavy dark matter in the Milky Way with velocity $\sim 0.001 c$ scattering elastically with nuclei or electrons, this assumption holds. 
Setting the arguments of the ``min" function equal to each other, the cross-section for which overburden scattering is saturated is manifestly $\sigmasos = n^{-2/3}$. 
Note that this saturation cross-section is independent of the length of the overburden, and depends only on its density.

\begin{figure}
\includegraphics[width=.49\textwidth]{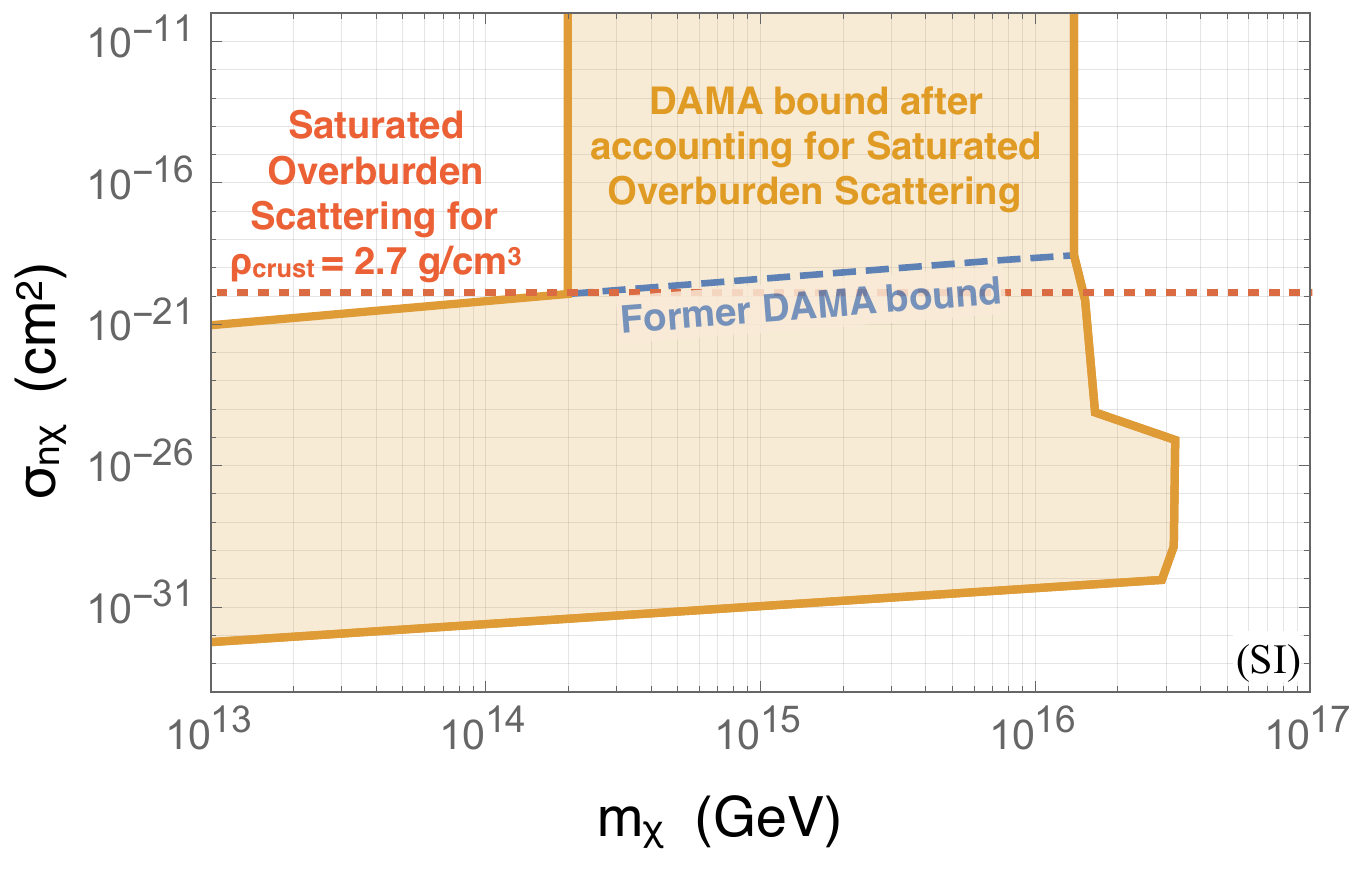} 
\caption{The bound from the \acro{DAMA} experiment on the spin-independent dark matter-nucleon scattering cross-section, before~\cite{Bernabei:1999ui} (dashed blue) and after (solid yellow) accounting for saturated overburden scattering in Earth with density $\rho_{\rm crust}$ = 2.7 g/cm$^3$. In the region above the dotted (red) line, dark matter scatters with every nucleus in the atmosphere and Earth's crust on its way to the detector. Therefore, increasing the dark matter-nucleon cross-section beyond this threshold does not increase the stopping power of the detector overburden. This excludes previously-allowed parameter space, up to a limit from the cooling of gas clouds in the Milky Way~\cite{Chivukula:1989cc}.}
\label{fig:sos}
\end{figure}

Typical dark matter direct detection experiments are conducted $\sim 1-2$ km underground, where the Earth's crust constitutes the main overburden. We find that for spin-independent scattering $\sigmasos$ = 1.3 $\times 10^{-20} \ {\rm cm}^2$, assuming a crust density of $2.7\1{g/cm^3}$ and a crust composition as given in Section \ref{sec:directionality}. Spin-dependent scatters occur mainly off $^{27}$Al, giving $\sigmasos \sim 10^{-18} \ {\rm cm}^2$. As illustrated in Figure~\ref{fig:sos}, the effect of saturating the overburden by scattering with all nuclei along the dark matter's path modifies prior bounds~\cite{Bernabei:1999ui}, extending the range of cross-sections probed to arbitrarily large values for very heavy dark matter.
We leave the re-visitation of other bounds vis-a-vis \acro{SOS} to future work, which include those from the \acro{RRS} balloon~\cite{Rich:1987st}, the \acro{XQC} calorimeter~\cite{Erickcek:2007jv,Mahdawi:2017cxz}, \acro{CRESST-I} Al$_2$O$_3$ experiments~\cite{Davis:2017noy,Kavanagh:2017cru}, as well as indirect bounds, $e.g.$ from Earth heating~\cite{Mack:2007xj}. We note that for a large enough nuclear scattering cross-section, dark matter prohibits the cooling of interstellar molecular gas clouds in the Milky Way~\cite{Chivukula:1989cc}, which however only becomes relevant for $\sigma_{\rm n \chi} \gtrsim 10^{-10}~{\rm cm^2} ~(\mDM/(10^{13}~{\rm GeV}))$ in the spin-independent case, where $\mdm$ is the dark matter mass.

\section{Dark Matter Scattering at High Multiplicity}\label{sec:multiplicity}

For the high dark matter masses considered here, the probed cross-sections are so large that particles may scatter with higher multiplicity than the customarily-assumed single scatter. Numerically, this can be understood by examining the number of dark matter particles found by a detector,
\beq
N_{\rm events} \sim \Phi \, \min[\tau,\; 1]\, ,   \label{eq:Nscatters}
\eeq
where $\Phi = (\rhodm/\mdm) \AD \vhalo \texp$ is the integrated flux of dark matter particles passing through the detector, $\AD$ the area of the detector, $\vhalo$ the average dark matter velocity, and $\texp$ the exposure (or observation) time. 
As dark matter passes through the detector, it will scatter a number of times roughly equal to the optical depth $\tau = n_{\rm det} \sigma\LD$ where $\LD$ and $n_{\rm det}$ are the detector length and number density of scattering sites in the detector.

Limits on $\sigma$ set by conventional searches are obtained in the single-scatter limit $\tau \ll 1$. Comparing $\Phi$ to $\tau$, it is clear that bounds on the cross-section weaken with increasing dark matter mass, $\sigma_{\rm bound} \propto \mdm^{-1}$. 
For high enough dark matter mass and hence cross-section, this scaling breaks down as $\tau \ra 1$, since transiting dark matter particles begin to scatter multiple times in the detector. In a typical single-scatter search, such events will have been cut away in the data analysis. As we will see, this breakdown occurs at a special point in ($\mdm,\sigmaT$) space where $\Phi \sim 1$ and $\tau \sim 1$. 
In the multiscatter limit, where all transiting dark matter scatters at least once ($\tau \gg 1$), it is the \emph{area} of the detectors which determines \acro{MIMP} sensitivity.
This can be compared to the single-scatter case, where the detector sensitivity is given by its volume or equivalently its total number of scattering sites.

If $\tau \gtrsim 1$, so that all dark matter particles scatter at least once in a detector, by setting the flux of through-going dark matter particles to one, $\Phi \sim 1$, we obtain the maximum mass to which an experiment is sensitive as $\mDMmax \sim \rhodm \AD \vhalo \texp$. For higher masses, the experiment lacks enough exposure to see any transiting dark matter. Typically, every generation of dark matter experiments increases the area $\AD$ of the detector as well as the exposure time $\texp$, thereby increasing $\mDMmax$. Thus for each new detector, there will be high-mass dark matter candidates that can only be uncovered with a multiscatter search. 
This {\em multiscatter frontier} encompasses a large portion of parameter space that needs be considered by such experiments.

\subsection{Prospects for Multiscatter Detection}\label{sec:prospects}

There is a substantial but presently unexplored region of dark matter parameter space for which dark matter scatters multiple times while traversing terrestrial detectors. A number of prior studies have considered dark matter scattering multiple times while transiting detectors~\cite{Rich:1987st,Starkman:1990nj,McGuire:1994pq,Bernabei:1999ui,Wandelt:2000ad,Albuquerque:2003ei,Zaharijas:2004jv,Erickcek:2007jv,Kouvaris:2014lpa,Davis:2017noy,Mahdawi:2017cxz,Kavanagh:2017cru,Hooper:2018bfw} and astronomical bodies~\cite{Starkman:1990nj,Mack:2007xj,Kavanagh:2016pyr,Bramante:2017xlb}.
Since our findings can be easily applied to a given experiment with adequate resolution and sensitivity to identify multiple scatters, we simply consider two example technologies for dark matter detection: liquid xenon time projection chambers and bubble chambers. Multiply interacting dark matter could be found by analyzing data already collected at these experiments, such as the Xe detectors \acro{XENON}\oldstylenums{1}\acro{T}~\cite{Aprile:2017iyp}, \acro{PandaX-II}~\cite{Cui:2017nnn}, and \acro{LUX}~\cite{Akerib:2016vxi}, and the bubble chamber experiment \acro{PICO}-\oldstylenums{60}~\cite{Amole:2017dex} containing C$_3$F$_8$. 
We also show the improvement in sensitivity at the future 50-ton detector \acro{DARWIN}~\cite{Aalbers:2016jon} and the future \acro{PICO} upgrade, \acro{PICO}-\oldstylenums{500}~\cite{pico500}. These experiments are poised to uncover \acro{MIMP}s in regions that are orders of magnitude beyond past efforts such as \acro{DAMA} (using NaI crystals)~\cite{Bernabei:1999ui}, and reanalyses of \acro{EDELWEISS} (Ge) and \acro{CDMS} (Ge, Si)~\cite{Albuquerque:2003ei}, which were able to search for dark matter up to $\mDM \lsim 10^{16}\1{GeV}$. 

For convenience, we define a \acro{MIMP} scattering threshold $\sigma_{\acro{MIMP}} = (\LD \ntarget)^{-1}$ as the cross-section at which the Poisson expectation value for the number of recoils from a single dark matter particle in a given detector $\tau = 1$. 
Given Poisson statistics, $\rm{Poiss}(\rm{multiplicity}, \; \tau)={({\rm e}^{-\tau} \tau^{\rm multiplicity})}/{({\rm multiplicity}!)}$, this happens when dark matter scatters at least twice during more than 25\% of its detector transits. We adopt canonical definitions for the dark matter-nucleon cross-section in the spin-independent (\acro{SI}) and spin-dependent (\acro{SD}) cases~\cite{Baudis:2012ig}. For a target with atomic number $A$, dark matter-nucleus reduced mass $\muTDM$, and dark matter-nucleon reduced mass $\munDM$, the \acro{SI} and \acro{SD} per-nucleon cross-sections in terms of the per-nucleus cross-section are given by
\bea
\nn \sigmaNDM^{({\rm \acro{SI}})} &=& \left(\frac{\munDM}{\muTDM}\right)^2 \left(\frac{1}{A^2}\right) \sigmaT \;\; \propto \;\; \left(\frac{1}{A^4}\right) \sigmaT~,\\
\nn \sigmaNDM^{({\rm \acro{SD}})} &=& \left(\frac{\munDM}{\muTDM}\right)^2 \left[\frac{4}{3} \frac{J_A + 1}{J_A} [a_p \langle S_p \rangle + a_n \langle S_n \rangle]^2\right]^{-1} \sigmaT \\ 
&\propto& \left(\frac{1}{A^2}\right) \sigmaT~,
\label{eq:pernucleonXS}
\eea
where the proportionality is in the $\mDM \gg \mtarget$ limit. For simplicity we will assume isospin-independent scattering and thus set $a_n=a_p=1$.

\begin{figure*}
  \includegraphics[width=.45\textwidth]{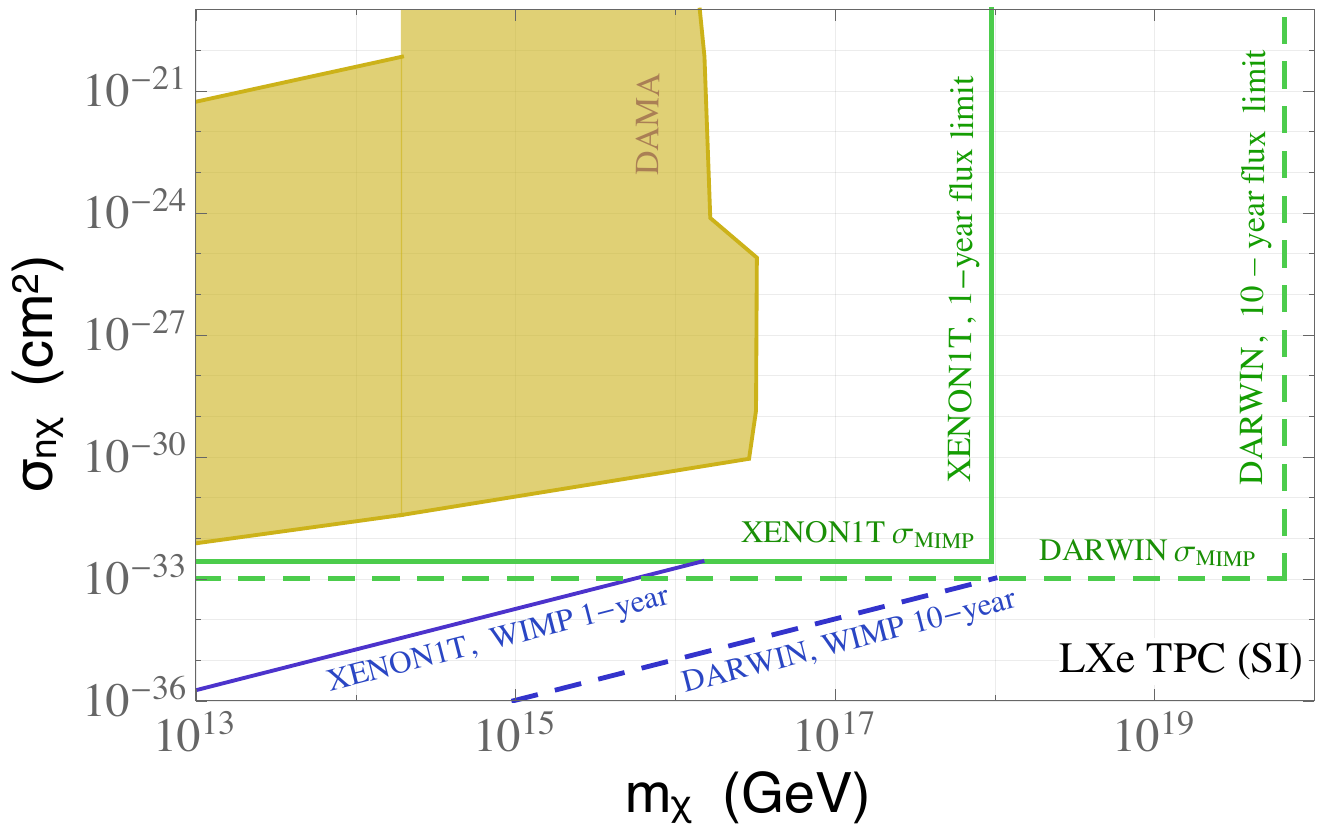} \quad \includegraphics[width=.45\textwidth]{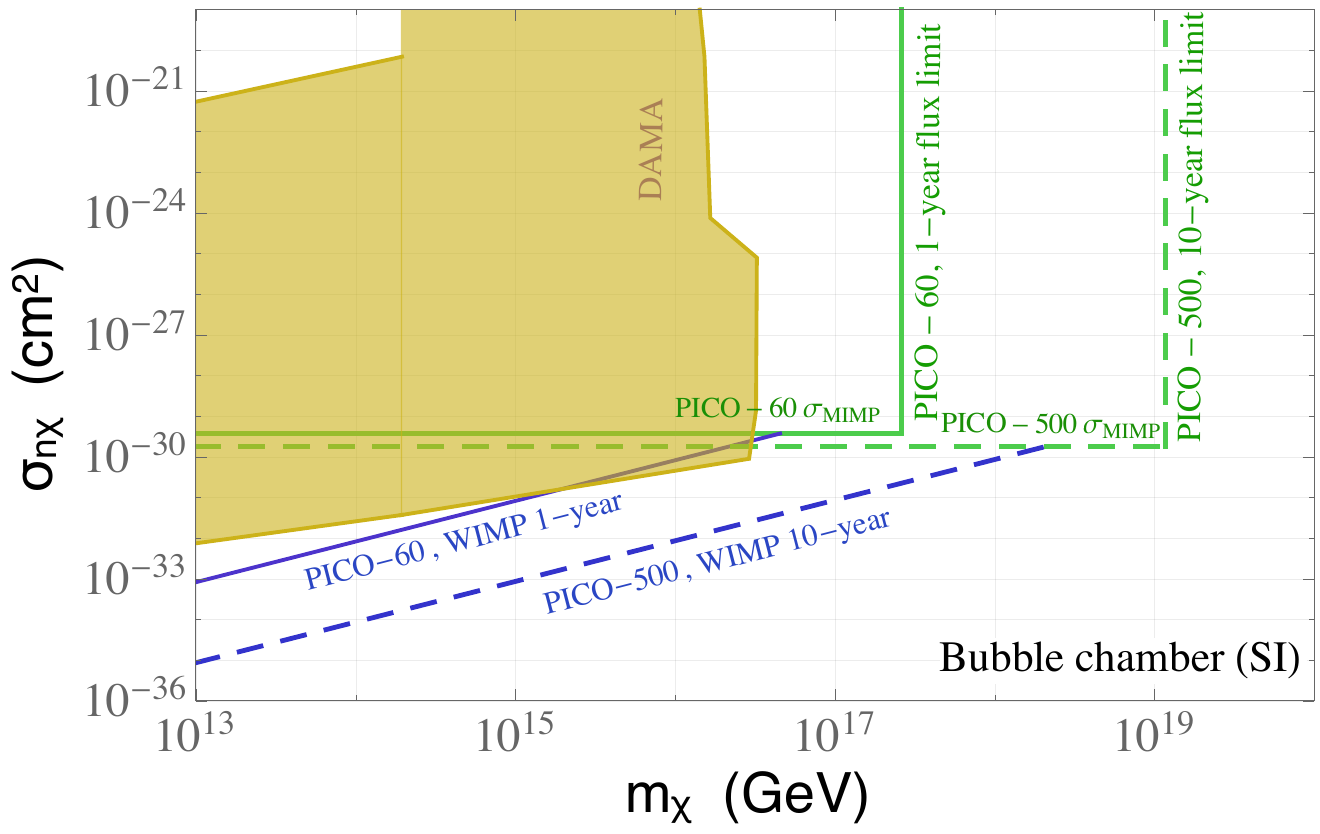} \\ 
  \includegraphics[width=.45\textwidth]{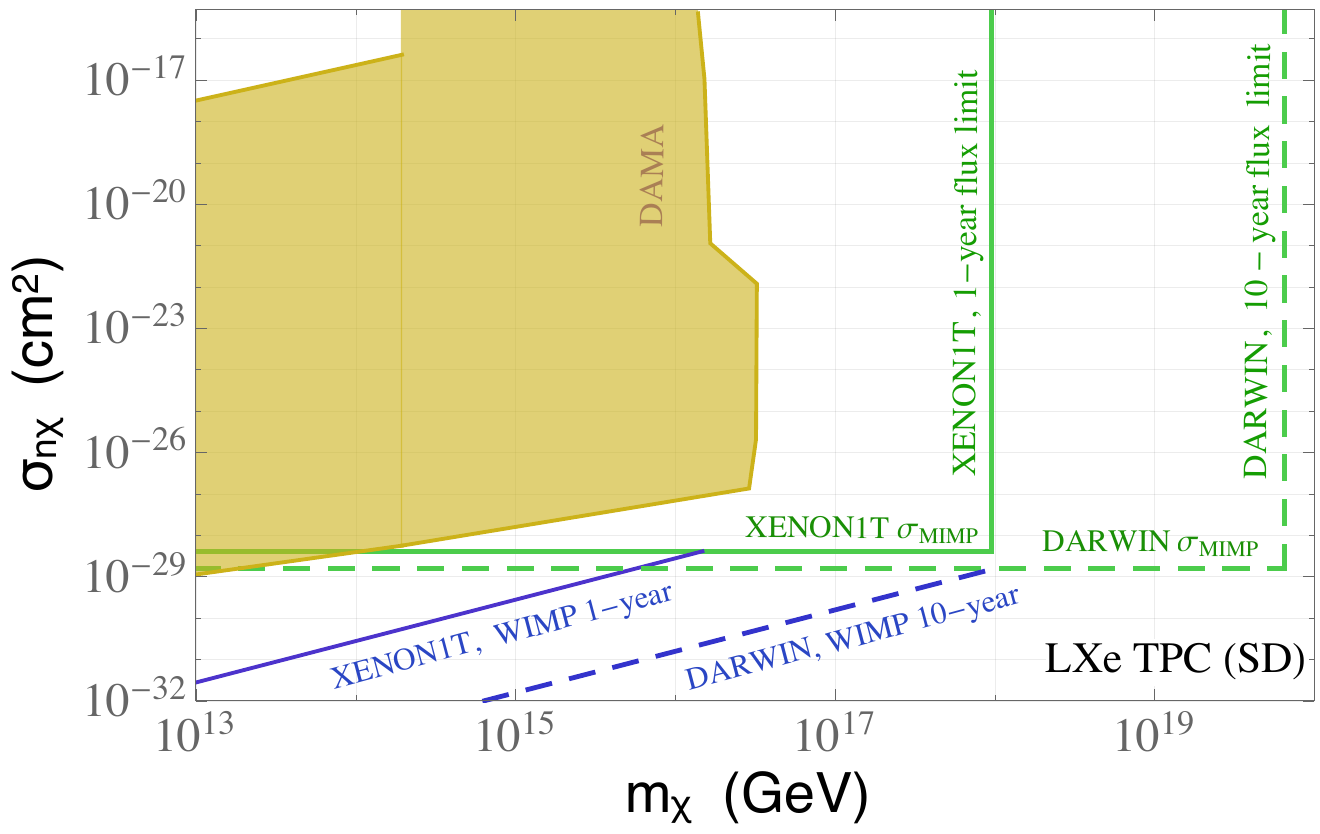} \quad \includegraphics[width=.45\textwidth]{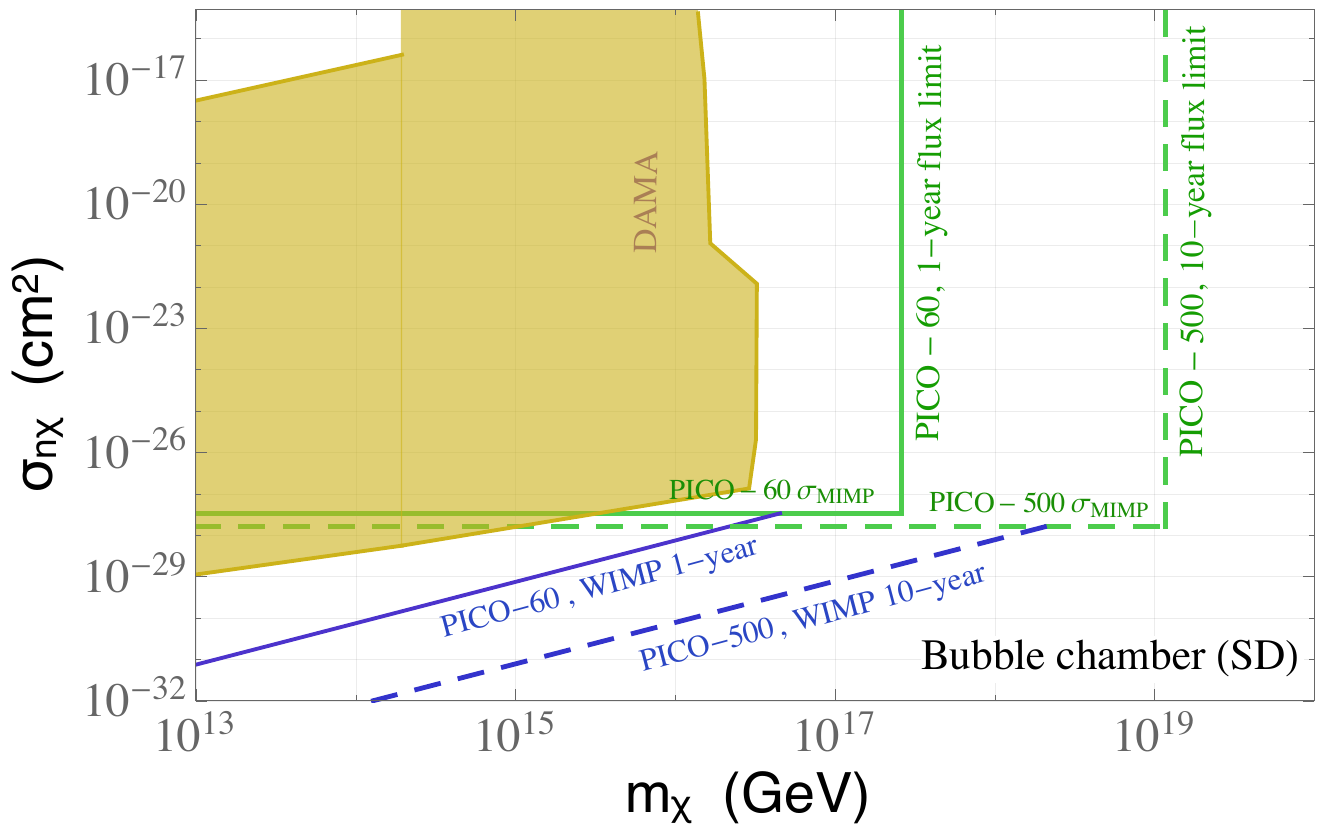} \\ 
  \caption{\label{fig:mimprospects} Prospects for probing dark matter scattering at high mass and high multiplicity, given the examples of liquid xenon or bubble chamber experiments. The (diagonal, blue) lines indicate the usual sensitivity from a zero-background single-scatter search. However, above the (horizontal, green) lines labeled $\sigmathresh$, more than 25\% of the dark matter traversing the detector scatters more than once, requiring a dedicated multiscatter analysis to probe this parameter range. The ultimate high-mass reach is given by the (vertical, green) lines from the requirement that at least 2.3~dark matter particles traverse the detector for the stated effective detector area and exposure time. 
For detectors such as \acro{DARWIN}, this limit can lie beyond the Planck mass $\sim 10^{19}\1{GeV}$.}
\end{figure*}

Figure~\ref{fig:mimprospects} shows the order-of-magnitude sensitivities achievable by \acro{MIMP} searches for events with high multiplicity, assuming a local dark matter density of $0.3\1{GeV/cm^3}$. 
In practice, regions covered by some searches where $\tau \gtrsim 1$ overlap due to Poisson fluctuations. Regions to the left of the solid green (dashed green) lines can be probed with 1-year (10-year) run-time. 
This mass reach is determined by requiring that the integrated flux $\Phi$ over the exposure time $\texp$ is at least 2.3 events (corresponding to the 90\%  confidence upper limit for zero observed events). 

Single-scatter sensitivities $\sigma_{\rm single}$ are shown for comparison, and are separated from \acro{MIMP} regions by horizontal green lines at the threshold cross-section $\sigmathresh$. These single-scatter sensitivities are taken from Refs.~\cite{Amole:2017dex,Aprile:2017iyp,Akerib:2017kat,Cui:2017nnn,Akerib:2016vxi} and are rescaled using the following parameters: for \{\acro{XENON}\oldstylenums{1}\acro{T}, \acro{DARWIN}, \acro{PICO}-\oldstylenums{60}, \acro{PICO}-\oldstylenums{500}\}, we simplify the detector as having an effective length $\LD = \{96\1{cm}$, $250\1{cm}, 50\1{cm}, 100\1{cm}$\}. We assume densities of $2.94\1{g/cm^3}$ for Xe and $1.36\1{g/cm^3}$ for C$_3$F$_8$, and isotopic abundances of 27\% $^{129}$Xe, 4\% $^{130}$Xe, 21\% $^{131}$Xe, 27\% $^{132}$Xe, 10\% $^{134}$Xe, and 9\% $^{136}$Xe, and 100\% of C$_3$F$_8$.
Following~\cite{Bednyakov:2004xq} we take $(J_A, \langle S_p \rangle, \langle S_n \rangle)=(\frac{1}{2}, 0.01, 0.3)$ for $^{129}$Xe, $(\frac{3}{2}, -0.04, -0.24)$ for $^{131}$Xe, and $(\frac{1}{2}, 0.48, -0.01)$ for $^{19}$F.

The relative behavior of these curves may be understood from their scalings with detector length, where for simplicity we here take a spherical geometry for the detector.
The multiscatter threshold cross-section then scales inversely with the effective diameter of the detector, $\sigmathresh \propto \LD^{-1}$. The maximum reachable mass scales with the effective target area and the exposure time, $\mDMmax \propto \LD^2 \texp$. 
As usual, the single-scatter cross-section sensitivity $\sigma_{\rm single} \propto (\LD^3 \texp)^{-1}$. We point out that depending on their actual exposure times, detectors such as \acro{XENONnT}~\cite{Aprile:2015uzo}, \acro{LZ}~\cite{Akerib:2018lyp}, \acro{DARWIN} and \acro{PICO}-\oldstylenums{500} are capable of probing dark matter masses of the order 10$^{19}$ GeV, $i.e.$ up to and beyond the Planck scale.

\subsection{Signatures at High Multiplicity}
\label{sec:features}

Perhaps the most salient feature of a super-heavy, multiply interacting, stable particle is that it will leave a mostly collinear track of nuclear recoils as it passes through a meter-scale detector. Comparing the maximum detector-frame scattering angle for \acro{MIMP}s, $\sin \alpha_{\rm max} = {\mtarget}/{\mx}$, with the maximum number of recoils in the detector $\ntarget^{1/3} \LD$, we find that in the $\mx \gg m_{\rm N}$ limit, the maximum total deflection angle of a dark matter particle is $\Omega_{\rm max} \lesssim n^{1/3} \LD \sin \alpha_{\rm max} $, or
\bea
\nn \Omega_{\rm max} & \lesssim & 1.7^\circ \left(\frac{\mtarget}{100~{\rm GeV}} \right) \left(\frac{10^{13}~{\rm GeV}}{\mx} \right) \times \\ 
& &  \left(\frac{\LD}{100~{\rm cm}} \right)\left(\frac{\ntarget}{10^{22}{\rm /cm^3 }} \right)^{1/3}.
\eea
Thus, the nuclear recoils of a transiting \acro{MIMP} are typically collinear, although for $\mx \lesssim 10^{13}\1{GeV}$, degree-level deflection along the length of a detector becomes plausible.
We remark that, since solar and atmospheric neutrinos will not leave tracks in detectors, the ``neutrino floor"~\cite{Billard:2013qya,Ruppin:2014bra} is less of a concern for \acro{MIMP} searches.

Detectors such as \acro{LUX}, \acro{PandaX-II} and \acro{XENON}\oldstylenums{1}\acro{T} are liquid xenon time projection chambers (\acro{TPC}s) where, for each interaction, two signals are observed with photomultiplier tubes: an $\Oc$(10 ns) pulse (``$S_1$") from scintillation in the scattering target, followed by an $\Oc$($\mu$s) pulse (``$S_2$") from electrolumiscence of electrons that have drifted into the gas above the target liquid. The drift time is $\Oc$(1 ms), allowing for clear separation of $S_1$ and $S_2$. In comparison to these timescales, dark matter transits a $1\1{m}$ detector length in $\sim5\1{\mu s}$. The relative strength of $S_1$ vs $S_2$ helps distinguish dark matter-induced nuclear recoils from electronic recoils from $\beta$ and $\gamma$ radiation, which comprises the main background.

A \acro{MIMP} transiting such a detector would produce multiple $S_1$'s and $S_2$'s, each characteristic of a nuclear recoil of relatively high energy in the range of 10's of keV. 
Whether the pulses would appear individually or merged is determined by the timing between successive scatters, which is typically the transit time divided by the number of recoils $\tau$. 
To understand the basic signatures let us neglect the fact that \acro{TPC}s have a much better resolution along their symmetry axis compared to the horizontal. 
Then, for $\tau \gsim 5$ ($\tau \gsim$ 500) the $S_2$ ($S_1$) pulses merge into elongated pulses $S'_2$ ($S'_1$). There are thus three qualitatively distinct \acro{MIMP} signatures: (1) a series of $S_1$s followed by a series of $S_2$s, for $1 \lsim \tau \lsim 5$; (2) a series of $S_1$s followed by a merged $S'_2$, for $5 \lsim \tau \lsim 500$; and (3) an elongated $S'_1$ followed by an $S'_2$, for $\tau \gsim 500 $, where the $S'_1$ and $S'_2$ will overlap at least partially. Backgrounds to \acro{MIMP} scattering will be exceedingly small. Signature~(1) at small multiplicity can be mimicked by the pile-up of individual single-scatter background events, which however will happen predominantly at the surface of the detector and thus can be fiducialized. Another potential background in this regime is from fast decays such as the $^{214}$BiPo coincidence which occurs in the $^{222}$Rn decay chain~\cite{Aprile:2017fhu,1748-0221-12-02-T02002}, but will be of little concern since the alpha-decay usually deposits much more energy than expected from \acro{MIMP}s. 
At intermediate multiplicity, radiogenic neutrons might mimic the expected \acro{MIMP} signature, but they do not usually travel at non-relativistic speeds nor scatter along a straight line. 
A background to elongated $S'_2$ events comes from instrumental sources of drifting electrons~\cite{Sorensen:2017ymt,Sorensen:2017kpl}, which will not usually conspire with multiple $S_1$ events to mimic the required \acro{MIMP} signature. 
At large multiplicity, tracks from through-going muons will display much shorter $S_1$ pulses than expected from \acro{MIMP}s, deposit much more energy than expected from dark matter, and can typically be vetoed by means of active shielding.

At \acro{PICO}-\oldstylenums{60}, the energy deposited by dark matter scattering off superheated C$_3$F$_8$ nucleates a bubble, identified visually with stereo cameras and acoustically with piezoelectric transducers. A \acro{MIMP} nucleates multiple bubbles during its $\sim$2.5 $\mu$s-long transit through the detector volume. The acoustics, sampled at a rate of 0.4 $\mu$s$^{-1}$, can identify at most 2.5/0.4 $\sim$ 6 individual bubble nucleations, whereas the cameras could image up to 100 bubbles. With the background from radioactive neutrons limited~\cite{Amole:2017dex}, the visual position reconstruction makes \acro{PICO} well-suited for a \acro{MIMP} search.

\section{determining the Mass, Cross-section, and local Density of MIMPs}
\label{sec:directionality}

In some regions of parameter space it is possible to determine the mass, scattering cross-section, and local density of \acro{MIMP}s using the directions of their arrival. 
To explain this, we must first detail our overburden calculations. 
The depletion of dark matter's kinetic energy after passing through a path $D$, along which lie nuclei with number densities $n_i$ and masses $m_i$ is
\begin{align}
\frac{E_f}{E_i} = \prod_{i}^{\rm nuclei} \left(1- z \beta_i \right)^{\tau_{\mathrm{od}, i}}~,
\label{eq:ovb}
\end{align}
where $E_i$ and $E_f$ are the respectively initial and final dark matter  kinetic energies and $z\beta_i = z\,{4 m_i \mx}/{(\mx + m_i)^2}$ is the fraction of kinetic energy lost in each scatter, where $z$ is a kinematic factor set to $1/2$ for evenly distributed center-of-mass scattering angles~\cite{Bramante:2017xlb}. 
To calculate this approximate upper bound on a detector's sensitivity to dark matter-nucleon scattering, we require that the final kinetic energy $E_f$ of a \acro{MIMP}, which travels at typical Galactic velocities of $10^{-3}c$, still has enough kinetic energy to deposit $1\1{keV}$ when recoiling against the target nucleus in an elastic collision.

\begin{figure}
  \includegraphics[width=.45\textwidth]{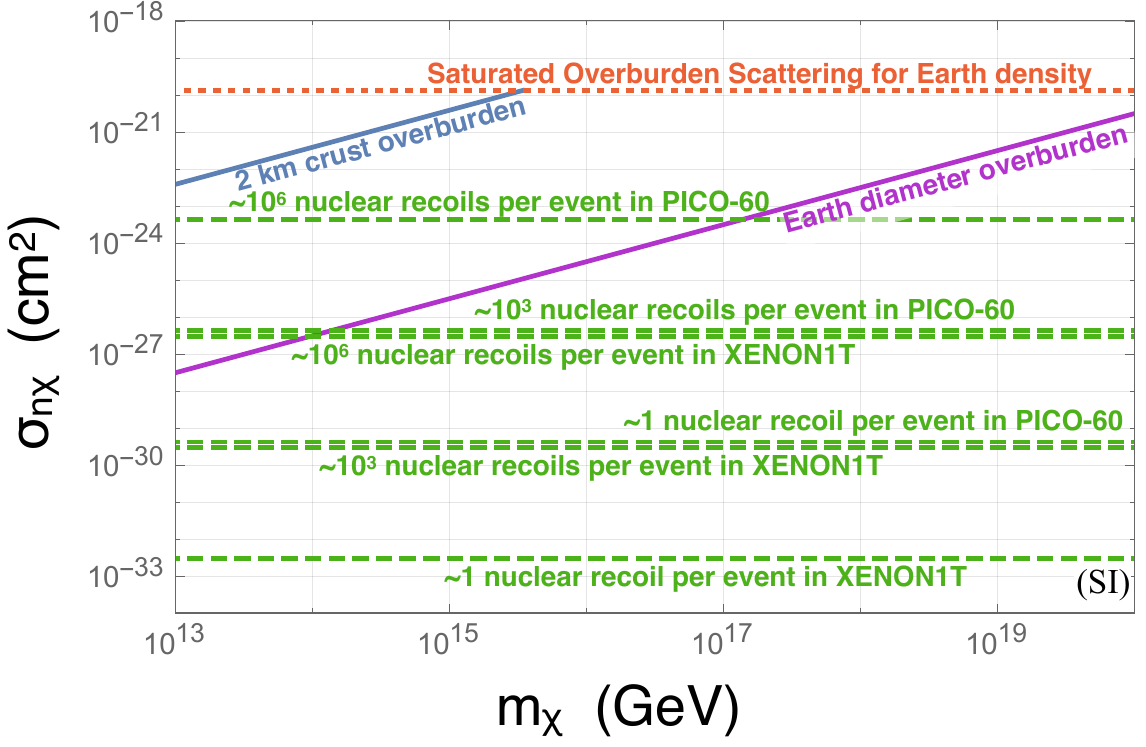}
  \caption{
The solid lines indicate the boundaries above which the \acro{MIMP} has lost too much of its kinetic energy propagating through the overburden, either from above through $2\1{km}$ of Earth crust or from below through $12756\1{km}$ of crust, mantle and core. In the parameter space between these lines, it is possible to use the acceptance cone of the detector to pinpoint the mass, scattering cross-section, and local density of \acro{MIMP}s. Dashed lines indicate the typical number of recoils from a single dark matter particle passing through \acro{Xenon}\oldstylenums{1}\acro{T} or \acro{PICO}-\oldstylenums{60}. The saturated overburden scattering for Earth with density $2.7~{\rm g/cm^3}$ is indicated with a dotted line.}
  \label{fig:directionaloverburden}
\end{figure}

To determine the number density of nuclei in the path of the dark matter as it transits either from above through the crust, or from below through the crust-mantle-core system, we model the interior of Earth using elemental abundances given in~\cite{Clarke1924,1980PNAS...77.6973M}. 
Specifically, we model the Earth as a sphere, and take Earth's crust to be $32\1{km}$ thick and composed of \{O, Si, Al, Fe, Ca, Na, K, Mg\} with fractions (by weight) \{0.467, 0.277, 0.08, 0.05, 0.03, 0.03, 0.03, 0.02\}. 
The mantle we take to have thickness $2867\1{km}$ and density 4.7 g/cm$^3$, composed of \{O,  Mg, Si, Fe,  Ca, Al\} with fractions \{0.448, 0.228, 0.215, 0.06, 0.02, 0.02\}. 
The core we take to have a radius of $3480\1{km}$ and density 12 g/cm$^3$, composed of \{Fe, Ni, S\} with fractions \{0.89, 0.06, 0.05\}. 
Figure~\ref{fig:directionaloverburden} shows the spin-independent dark matter-nucleon cross-sections above which the dark matter's kinetic energy is diminished so much that it becomes undetectable, for a path through $2\1{km}$ of the crust or for the opposing radial path through the Earth, passing through $62\1{km}$ of crust, $5734\1{km}$ of mantle and $6960\1{km}$ of core. Spin-dependent scattering occurs off Al, Na, Ca, and $^{29}$Si, with nuclear spin expectation values ($\langle S_p \rangle, \langle S_n \rangle$) given in~\cite{Bednyakov:2004xq}.

With the overburden specified, we now show how \acro{MIMP} directionality can be used to determine the mass, cross-section, and local density of \acro{MIMP}s with a detection in just a single detector. As a preliminary caution, we note that while typical directional dark matter detection effects rely on a spatial anisotropy of incoming dark matter velocities (the dark matter velocity towards Earth is peaked in a direction pointing toward Cygnus), the effect we introduce here does \emph{not} depend on dark matter velocity anisotropy. This directional effect would occur even if the dark matter velocity were perfectly isotropic at the surface of Earth, and arises simply because the overburden above and below a direct detection experiment is different.

The directional analysis we propose proceeds as follows: first, the number of recoils from a single \acro{MIMP} in the detector gives a measurement of the optical depth $\tau = n_{\rm det} \sigma_{T \chi} \LD$, where the detector length $\LD$ and nuclear density $n_{\rm det}$ are known. Thus the \acro{MIMP} scattering cross-section can be determined directly, and independently of the local dark matter density, by measuring the number of recoils per \acro{MIMP} event along the length of a detector. 
Next, for parameter space between the two diagonal overburden lines in Figure~\ref{fig:directionaloverburden}, some fraction of the \acro{MIMP}s will not arrive at the detector. 
In fact, the arriving flux will have an angular dependence, since upward-going \acro{MIMP}s will pass through more Earth overburden. 
More precisely, the \acro{MIMP} mass and cross-section uniquely determine the ``angle of acceptance" at a direct detection experiment, defined as the angle between a vector pointing in the normal direction from the detector to the Earth's surface, and a vector pointing from the detector towards where the dark matter flux through the Earth is halved. 
This halving of flux will occur near an angle where all dark matter particles coming up through the Earth are blocked, which depends on dark matter's mass and velocity distribution; see Equation~\eqref{eq:ovb}. 
After appropriately modeling the dark matter velocity distribution and the Earth's interior, it would be possible, using Equation~\eqref{eq:ovb}, to pinpoint the dark matter mass. (This is in contrast to usual single-scatter \acro{WIMP} analyses where at high masses the spectral shape carries no information about the \acro{WIMP} mass.)
With the \acro{MIMP} flux measured by the experiment ($\Phi = (\rhodm/\mdm) \AD \vhalo \texp$) this is enough to determine the local \acro{MIMP} mass density.

A more detailed simulation would also incorporate the orientation of the dark matter speed, which is roughly maximal towards the Cygnus constellation, relative to the dark matter detector. Intriguingly, since the acceptance angle depends slightly on the dark matter velocity distribution, it is larger when the detector is so oriented that the higher-velocity \acro{MIMP}s from Cygnus are traveling through more of Earth. A full analysis thus requires keeping track of the detector's varying orientation with respect to the dark matter wind due to Earth's rotation and orbit around the Sun, as well as the arrival times of observed \acro{MIMP} events.

\section{Outlook}
\label{sec:discs}

Super-massive relic particles with per-nucleon cross-sections in the $10^{-40} - 10^{-20}~{\rm cm^2}$ range are motivated by grand unified, supersymmetric, and other theories which contain stable heavy particles charged under Standard Model gauge symmetries. 
We have demonstrated that multiply scattering massive particles, with masses in the $10^{10}-10^{20}$ GeV range, require dedicated analyses, and may be discovered in data already collected by dark matter direct detection experiments, along with future experimental searches. 
As explained in Section \ref{sec:multiplicity}, it is no coincidence that  the multiscatter frontier opens up just as the flux limit is approached for a given detector: as dark matter particles become rarefied with increasing mass, a detector has to be able to interact with every transiting particle for a meaningful signal. 
For cross-sections larger than this, dark matter will scatter multiple times in the detector.

The new phenomenology presented in this article reveals many avenues for future research. Previously published limits on dark matter at large cross-sections must be re-derived after taking saturated overburden scattering into account. The sensitivity of dark matter detectors to leptophillic \acro{MIMP}s scattering multiple times with target electrons must be explored. A directional analysis for validating \acro{MIMP} signals at direct detection experiments incorporating a more detailed model of the Earth's interior can be undertaken. Dropping the assumption of simple spherical detectors and taking into account the direction of drift in liquid xenon \acro{TPC}s will result in diurnal modulation signatures. \acro{MIMP} searches at detectors such as Borexino~\cite{Alimonti:2000xc}, \acro{SNO}+~\cite{Andringa:2015tza}, \acro{JUNO}~\cite{Li:2014qca}, and \acro{MATHUSLA}~\cite{Chou:2016lxi} should be investigated. Finally, the links between grand unified theories with stable colored and electroweak states, the cosmological production of the same, and terrestrial searches for \acro{MIMP}s should be sought out. Altogether, the search for dark matter in multiple-scatter events has the potential to revolutionize our understanding of fundamental physics and the early universe.

\section*{Acknowledgments}

We thank
 Jimmy Bramante,
 Mark Chen,
 Ken Clark,
 Adrienne Erickcek,
 Jonathan Feng,
 Gilles Gerbier,
 Guillaume Giroux,
 Bradley Kavanagh,
 Graham Kribs,
 Jason Kumar,
 Adam Martin,
 Sam McDermott,
 Tony Noble,
 Maxim Pospelov,
 Jessie Shelton,
 Louis Strigari,
 Aaron Vincent,
 and
Alex Wright
for useful conversations. 
We thank the Perimeter Institute for Theoretical Physics, where this work was initiated. Research at Perimeter Institute is supported by the Government of Canada through Industry Canada and by the Province of Ontario through the Ministry of Economic Development \& Innovation. 
J.~B.~ and B.~B.~ acknowledge the support of the Natural Sciences and Engineering Research Council of Canada. 
J.~B.~thanks the Aspen Center for Physics, which is supported by National Science Foundation grant PHY-1066293. 
NSF also supports R.F.L. through grant PHY-1719271 and 
N.R. through grant PHY-1417118.

\bibliography{christmas}

\end{document}